\documentclass[aps,prd,reprint,onecolumn,nofootinbib,longbibliography,12pt,superscriptaddress]{revtex4-2}
\usepackage{textcomp}
\DeclareUnicodeCharacter{2212}{\textminus}

\usepackage{graphicx}
\usepackage{dcolumn}
\usepackage{bm}
\usepackage{amsmath}
\usepackage{amssymb}
\usepackage{booktabs}
\usepackage{setspace}
\usepackage{siunitx}
\usepackage{caption}
\usepackage{subcaption}
\usepackage{microtype}
\usepackage[colorlinks=true, linkcolor=blue, citecolor=blue, urlcolor=blue]{hyperref}

\begin{document}

\title{Early- and Late-Time Modifications to \texorpdfstring{$\Lambda$}{Lambda}CDM: Implications for the Hubble Tension}

\author{Rahul Dhyani}
\email{rd947@snu.edu.in}
\affiliation{%
Shiv Nadar Institution of Eminence (Deemed to be University), Tehsil Dadri, Gautam Buddha Nagar, Uttar Pradesh 201314, India%
}%

\author{Purba Mukherjee}
\email{pdf.pmukherjee@jmi.ac.in}
\affiliation{%
Centre for Theoretical Physics, Jamia Millia Islamia, New Delhi 110025, India%
}%
\affiliation{%
Korea Astronomy and Space Science Institute (KASI), Daejeon 34055, Republic of Korea%
}%

\author{Arindam Chatterjee}
\email{arindam.chatterjee@snu.edu.in}
\affiliation{%
Shiv Nadar Institution of Eminence (Deemed to be University), Tehsil Dadri, Gautam Buddha Nagar, Uttar Pradesh 201314, India%
}%

\author{Anjan A Sen}
\email{aasen@jmi.ac.in}
\affiliation{%
Centre for Theoretical Physics, Jamia Millia Islamia, New Delhi 110025, India%
}%

\date{\today}
\begin{abstract}
We investigate an extension of $\Lambda$CDM in which a 
fraction of cold Dark Matter (DM) decays into invisible dark radiation (DR) 
around the radiation--matter equality epoch, together with a non-standard dark 
energy (DE) equation of state characterized by $w_0$. The decaying DM 
component modifies the early expansion history and reduces the sound 
horizon at baryon drag, while the DE alters the expansion rate 
at the late times. A comprehensive analysis combining \texttt{Planck 2018+ACT 
DR6+DESI DR2+CMB lensing} datasets has been carried out to explore the 
viability of this framework in addressing the $H_0$ tension. This model 
yields a Hubble constant of $H_0 = 69.83 \pm 0.98~\mathrm{km\,s^{-1}\,Mpc^{-1}}$, 
reducing the discrepancy with SH0ES measurement to ${\sim}2.2\sigma$ and local distance network  measurement 
(H0DN) to ${\sim}2.9\sigma$. Further, 
considering \texttt{SH0ES} and \texttt{Pantheon+}, the inferred value of 
the Hubble constant becomes $H_0 = 70.20 \pm  
0.66~\mathrm{km\,s^{-1}\,Mpc^{-1}}$. The Bayesian evidence 
suggests that this framework offers a fit to the relevant cosmological datasets at a statistically similar level as $\Lambda$CDM. It is observed  that correlated early- and late-time modifications to the cosmological expansion history provide a more effective route to reducing the $H_0$ tension than either class of modification alone.
\end{abstract}
\maketitle
\baselineskip=0.8\baselineskip

\section{Introduction}
\label{sec:intro}

The simplest incarnation of the standard model of cosmology, $\Lambda$CDM, 
where $\Lambda$ and CDM stand for cosmological constant and Cold Dark Matter, 
respectively, has been successful in light of a wide range of cosmological 
observations. These include observations probing the early and late stages 
of the evolution of our Universe, notably the Cosmic Microwave Background 
Radiation (CMBR) anisotropies \citep{Planck2018,AtacamaCosmologyTelescope:2025nti}, 
as well as the large-scale structure (LSS) mapped by modern galaxy surveys\citep{SDSS:2014iwm,eBOSS:2020yzd,DES:2024jxu,DESI:2024uvr,DESI2024,DESI:2025fii,DES:2026fyc}. Despite its remarkable success, 
increasing observational precision has revealed persistent tensions among independent 
cosmological probes. The most prominent among these are the discrepancies between 
early- and late-Universe determinations of the Hubble constant $H_0$ 
\citep{Riess2022,Planck2018,AtacamaCosmologyTelescope:2025nti} and the amplitude 
of matter clustering inferred from weak-lensing and galaxy surveys compared to 
the CMB predictions, commonly expressed as the $S_8$ tension \cite{DiValentino:2020vvd,Pantos:2026koc}. 
In particular, the discrepancy in the $H_0$, is significant and persists at the $(5-7)\sigma$ 
level  \citep{Planck2018,DESI2024,AtacamaCosmologyTelescope:2025nti,DESI:2025fii,
Riess2022,Verde:2023lmm,H0DN:2025lyy}, see also, e.g., \citep{Verde:2019ivm,
DiValentino:2020zio,Perivolaropoulos:2021jda}. Further, while results from the KiDS 
Legacy Survey indicate that the previously reported $S_8$ tension is no longer 
significant \citep{Stolzner:2025htz}, recent DES-Y6 \cite{DES:2026mkc} 
results find that the tension persists at $\sim 2.8\sigma$ \cite{DES:2026fyc}. 

From the CMBR observations, assuming $\Lambda$CDM, \texttt{Planck} collaboration 
obtains $H_0 = 67.4 \pm 0.5  \, \mathrm{km\,s^{-1}\,Mpc^{-1}}$ \citep{Planck2018}, 
while combining Planck and Atacama Cosmology Telescope \texttt{(ACT)} observations 
$H_0 = 67.64 \pm 0.5 \, \mathrm{km\,s^{-1}\,Mpc^{-1}}$ \cite{ACT2025}. The local 
distance ladder measurement of $H_0$ via Cepheid-calibrated SNeIa from 
\texttt{SH0ES} yields $H_0 = 73.04 \pm 1.04 \, \mathrm{km\,s^{-1}\,Mpc^{-1}}$ 
\citep{Riess2022}. The Local Distance Network (\texttt{H0DN}) collaboration 
has reported a local measurement of the Hubble constant combining multiple 
independent distance indicators, and obtains $H_0 = 73.50 \pm 0.81,
\mathrm{km,s^{-1},Mpc^{-1}}$\citep{H0DN:2025lyy}. The estimation of $H_0$ differs 
from the joint \texttt{Planck+SPT-3G+ACT (DR6)} constraint at the level of 
$7.1\sigma$~\citep{SPT-3G:2025bzu}, and from the 
\texttt{Planck+ACT (DR6)+DESI(BAO) DR2} combination at approximately $6\sigma$. Such a 
persistent and high-significance disagreement, termed as the Hubble tension, has attracted 
substantial attention in the literature \cite{DiValentino:2016hlg,Karwal:2016vyq,Freedman:2017yms,
Feeney:2017sgx,DiValentino:2021izs}. 

The proposed resolutions may be broadly classified as follows. The first category includes 
modifications within the standard cosmological framework which alter processes, such as the 
reionization history \citep{Chatterjee:2021ygm,Allali:2025wwi} or recombination \footnote{In this context, it may 
be noted that as compared to the \texttt{Planck} CMB determination $\tau_{\rm reio} = 0.0544 
\pm 0.0073$ \citep{Planck2018}, recent results based on \texttt{JWST} observations indicate 
a higher reionization optical depth, possibly reaching $\tau_{\rm reio} \gtrsim 0.07$ 
\citep{Munoz:2024fas}.}\citep{Liu:2019awo,Rashkovetskyi:2021rwg,Mirpoorian:2024fka,Lynch:2024hzh,Jedamzik:2025cax}.
The second category involves modification of the dark sector and/or the gravity sector 
beyond the simplest realization by introducing new degrees of freedom, novel interactions, 
and/or dynamics, modifying the evolution of the early Universe \citep{Zhang:2007zzh,Chudaykin:2016yfk,DiValentino:2017zyq,Odintsov:2017qif,Nunes:2018xbm,Mortsell:2018mfj,Poulin:2018cxd,Odintsov:2018qug,Knox:2019rjx, Pandey:2019plg,Wang:2020zfv,Ballardini:2020iws,Ballesteros:2020sik,Zumalacarregui:2020cjh,Clark:2020miy,Haridasu:2020xaa,Wang:2020dsc,DiValentino:2021izs,Schoneberg:2021qvd,Hubert:2021khy,Sen:2021wld,Hazra:2022rdl,Kamionkowski:2022pkx,Bucko:2022kss,Gomez-Valent:2023uof,Adil:2023exv,Tiwari:2023jle,Poulin:2023lkg,Efstathiou:2023fbn,Seto:2024cgo,Poulin:2024ken,Toda:2024ncp,CosmoVerseNetwork:2025alb,Teixeira:2025czm,Mukherjee:2025myk,Cheng:2025lod,Shah:2025ayl}. Such scenarios aim to reconcile early- and late-time 
determinations of $H_0$ by altering the pre- or post-recombination dynamics, while 
remaining consistent with other cosmological observations. Further, various systematic 
uncertainties have been explored as possible sources \citep{Mortsell:2021nzg,Mortsell:2021tcx,
Freedman:2021ahq}.

It has been noted that an increase in $H_0$ without modifying the (comoving) sound 
horizon at baryon drag $r_{\rm drag}$ leads to tensions with BAO and Hubble parameter estimations 
from type-Ia supernovae measurements, as the (comoving) angular diameter distance 
$(D(z))$ scales inversely with $H_0$. As BAO measurements tightly constrain 
the combination $r_{\rm drag}/D(z)$, therefore, extensions of $\Lambda$CDM which only affect 
the evolution of the Universe at late times has been less successful in resolving the 
Hubble tension. Further, as the sound horizon $r_{\rm drag} = \int_{z_d}^{\infty} \frac{c_s(z)}
{H(z)} \, dz$ scales inversely with the Hubble parameter $H(z)$, an increase in the 
early expansion rate $H(z)$ due to the additional energy component leads to a decrease 
in $r_{\rm drag}$ \cite{Jedamzik:2020zmd,Evslin:2017qdn}.\footnote{
For example, Early Dark Energy 
(EDE) models, in which a scalar field temporarily contributes a non-negligible fraction 
of the total energy density before the recombination epoch and then rapidly dilutes, 
thereby increasing $H(z)$ at early times without significantly affecting late-time 
cosmology. BAO measurements tightly constrain the combination $r_{\rm drag}/D(z)$, and a 
smaller $r_{\rm drag}$ requires corresponding shifts in distance measures that are not always 
favored by low-redshift data \citep{Poulin:2018cxd,Kamionkowski:2022pkx,Poulin:2023lkg}.}

The present study considers an amalgamation of the early-time and late-time 
modifications of the evolution of the Universe. In particular, it incorporates 
a decaying fraction of DM, parametrized by $f_{\rm dcdm}$, with a decay width 
$\Gamma_{\rm dcdm}$, such that $\Gamma_{\rm dcdm}^{-1} \simeq H_{EQ}$, where $H_{EQ}$ 
denotes the Hubble parameter at the radiation--matter equality epoch and the subscript 
``dcdm" stands for decaying DM. For the 
late-time modification, the Dark Energy (DE) equation of state parameter $w_0$, 
has been introduced. Further, the Chevallier--Polarski--Linder (CPL) parametrization  
with $w(a) =w_0 +w_a\, (1-a)$, where ($w_0, w_a$) are the model parameters 
\citep{CPL2001,Linder2003} has also been considered. 

Each of these extensions of $\Lambda$CDM have been widely studied in the literature. 
In particular, in the context of decaying DM, two possibilities have been considered 
in the literature: a one-parameter extension characterized solely by the decay rate 
$\Gamma_{\rm dcdm}$, and a two-parameter extension in which a fraction $f_{\rm dcdm}$ 
of the CDM decays with rate $\Gamma_{\rm dcdm}$ into (relativistic) dark radiation (DR) 
and/or into warm dark matter (WDM) \citep{Berezhiani:2015yta,Chudaykin:2016xhx,Xiao:2019ccl,Nygaard:2020sow,Blinov:2020uvz,FrancoAbellan:2021sxk,Holm:2022eqq,Alvi:2022aam,Anchordoqui:2022gmw,Davari:2022uwd,Simon:2022ftd,Simon:2022hpr,Gambini:2022zom,Bucko:2024izb,Zhou:2025ikl,Juarez-Jimenez:2025ktm}. 
The resolution of the $H_0$ tension depends on the lifetime, given by the inverse 
of the decay width ($\Gamma_{\rm dcdm}^{-1}$), and on the fraction of decaying DM($f_{\rm dcdm}$). 
Such a decay process in the early Universe leads to modifications in both the background 
energy densities and the evolution of cosmological perturbations. As for modifications 
of the DE equation of state, this is generally parameterized either by a constant value, 
$w=w_0$, or by a time-varying function $w(a)$, which, in Chevallier--Polarski--Linder 
(CPL) parametrization takes the following form $w(a) =w_0 +w_a\, (1-a)$, where ($w_0, w_a$) 
are the model parameters \citep{CPL2001,Linder2003}.  
Such parametrizations can provide an effective phenomenological description 
of physical DE models \cite{Scherrer:2015tra} and successfully reproduce the 
observable signatures of quintessence scenarios \cite{Tsujikawa:2013fta}. It has been 
suggested that, while each of these extensions individually leads 
to an increase in $H_0$ to around  $H_0 \simeq 69 \, \mathrm{km\,s^{-1}\,Mpc^{-1}}$, these 
generally fall short of fully resolving the tension \citep{Jedamzik:2020zmd,Vagnozzi:2023nrq}.  This motivates the approach explored in this article, where fractional 
DCDM with $\Gamma_{\rm dcdm}^{-1} \simeq H_{EQ}$ together with modification of the equation of state 
for DE has been considered. In light of recent cosmological observations from 
\texttt{ACT DR6}~\citep{ACT2025} and \texttt{DESI DR2}~\cite{DESI:2025zgx}, in addition to \texttt{Planck}~\cite{Planck2018}, \texttt{SH0ES}~\cite{Riess2022} and \texttt{Pantheon+}~\cite{Brout:2022vxf} a comprehensive analysis  has been performed to explore the viability 
of this framework in addressing the $H_0$ tension.

This paper is organized as follows. In Sec.~\ref{sec:model}, we present the theoretical 
framework, describing the fractional decaying cold dark matter ($f {\rm DCDM}$) sector and DE Parametrization. Our main results, including parameter constraints, posterior distributions, 
impacts on the Hubble constant $H_0$ tension, and comparisons with $\Lambda$CDM, are presented 
and discussed in Sec.~\ref{sec:results}. Finally, Sec.~\ref{sec:conclusion} provides a 
summary of the key findings, discusses implications for cosmology, and outlines directions 
for future work.

\section{The Framework}
\label{sec:model}

In this section, we describe the extended cosmological framework that is considered 
for the present study. The base $\Lambda$CDM, has been extended to include three additional 
parameters in the CDM and DE sectors. In the CDM sector, it is assumed that a fraction 
$f_{\rm dcdm}$ of the total CDM abundance decays into invisible dark radiation with a decay 
width $\Gamma_{\rm dcdm}$. For DE, the equation of state has been parametrized to account 
for non-trivial evolution of the respective energy density. 

The combination of these two sectors allows correlated adjustments to early- and late-universe 
observables, offering a more flexible extension of $\Lambda$CDM than either modification 
allows for in isolation.

\subsection{\texorpdfstring{Fractional Decaying Cold Dark Matter ($f \mathrm{DCDM}$)}
{Fractional Decaying Cold Dark Matter (fDCDM)}}
\label{subsec:dcdm}

In the present case, we consider a fraction of the DM which decays into invisible relativistic 
species (denoted as DCDM), hereafter referred to as dark radiation (DR) \citep{Poulin:2016nat,
Audren2014,Xiao:2019ccl,Nygaard:2020sow,Alvi:2022aam}. The decay is characterized by a constant 
decay width $\Gamma_{\rm dcdm}$. The fractional abundance of the decaying component at a time 
$t_i \ll \Gamma_{\rm dcdm}^{-1}$ in the cosmic comoving frame is denoted by $f_{\rm dcdm}$, with the 
remaining fraction $(1 - f_{\rm dcdm})$ consists of stable and non-interacting CDM (dubbed 
as SCDM). Since the decay produces DR, there are no constraints from electromagnetic energy 
injection and CMB spectral distortions. Thus, for 
$ t \ll 1/\Gamma_{\rm dcdm}$ the entire CDM abundance comprises of DCDM and SCDM, while for  $ t \gg 1/\Gamma_{\rm dcdm}$ 
only SCDM contributes to the DM energy density. In the following discussion, this CDM sector 
will be referred to as $f$DCDM.

The decay process leads to energy transfer from DM to DR. The total energy--momentum tensor 
$T^{\mu\nu}_{\rm T} = T^{\mu\nu}_{\rm dr}+T^{\mu\nu}_{\rm dcdm}$, where the subscripts $\rm dr$ 
and $\rm dcdm$ stand for contributions from DR and DCDM components, respectively, is covariantly 
conserved, i.e. $\nabla_\mu T^{\mu\nu}_{\rm T} = 0$. Further, for $T^{\mu\nu}_{\rm dr}, 
~T^{\mu\nu}_{\rm dcdm}$, we have, 
\begin{equation}
\nabla_\mu T^{\mu\nu}_{\rm dcdm} = -Q^\nu, \qquad
\nabla_\mu T^{\mu\nu}_{\rm dr} = Q^\nu,
\end{equation}
where the effect of the energy-momentum transfer from DCDM to DR is described by the four-vector 
$Q^\nu$ as follows, 
\begin{equation}
Q^\nu = a \Gamma_{\rm dcdm} \rho_{\rm dcdm} u^\nu_{\rm dcdm},
\end{equation}
with $u^\nu_{\rm dcdm}$ denoting the four-velocity of the DCDM fluid. The number of DCDM particles decaying per unit volume per unit comoving time is proportional to the decay width multiplied by the respective number density. Consequently, the rate of energy transfer per unit volume (in the comoving frame) to DR because of DCDM decay is proportional to this number density particles multiplied by the rest mass of the DCDM, and thus, $\Gamma_{\rm dcdm} \rho_{\rm dcdm}$. The four-velocity of DCDM ($u^{\nu}_{\rm dcdm}$) in the comoving frame, which is also the rest-frame of DCDM (i.e., $u^{0}_{\rm dcdm} =1$), ensures the covariance of the above equation. This choice corresponds 
to decay in the rest frame of the parent particle. The resulting evolution equations for the 
background energy densities of DCDM ($\rho_{\rm dcdm}$) and DR ($\rho_{\rm dr}$) are
\begin{align}
\rho'_{\rm dcdm} + 3\mathcal{H} \rho_{\rm dcdm} &= -a \Gamma_{\rm dcdm} \rho_{\rm dcdm},
\label{eq:dcdm-bg} \\
\rho'_{\rm dr} + 4\mathcal{H} \rho_{\rm dr} &= +a \Gamma_{\rm dcdm} \rho_{\rm dcdm},
\label{eq:dr-bg}
\end{align}
where prime denotes derivative with respect to conformal time, $\mathcal{H} \equiv a'/a$ is 
the conformal Hubble parameter, and $a$ is the scale factor. 

As described in Sec.\ref{sec:intro}, $f$DCDM framework is parameterized by the initial fractional 
abundance $f_{\rm dcdm}$, and the decay width $\Gamma_{\rm dcdm}$ of the DCDM. The first parameter is 
defined as the relative abundance of DCDM, and is given by, 
\begin{equation}
f_{\rm dcdm} = \frac{\rho_{\rm dcdm}(t_{i})}{\rho_{\rm dcdm}(t_{i})+\rho_{\rm scdm}(t_{i})}=
\frac{\Omega_{\rm dcdm}(t_{i})}{\Omega_{\rm dcdm}(t_{i})+\Omega_{\rm scdm}(t_{i})},
\end{equation}
where $\Omega_{\rm cdm} \equiv \Omega_{\rm dcdm}(t_{i}) + \Omega_{\rm scdm}(t_{i})$ 
is the total CDM contribution. By construction, $f_{\rm dcdm} \in [0,\,1]$, with $f_{\rm dcdm} = 0$ 
corresponding to the standard $\Lambda$CDM scenario, and $f_{\rm dcdm} = 1$ corresponding 
to the case in which the entire CDM abundance is unstable.

The conversion of CDM into DR has several consequences, which depend on the 
timescale $\Gamma_{\rm dcdm}^{-1}$. At $t \simeq \Gamma_{\rm dcdm}^{-1}$, $\rho_{\rm dr}(t)$ increases, leading to 
a small increment of the sound speed $c_s$ in the same epoch. However, a larger matter 
abundance (in comparison with radiation) for $t > \Gamma_{\rm dcdm}^{-1}$ leads to an enhancement 
in the Hubble parameter $H(z)$ at the corresponding redshifts $z$, and, therefore, the 
integral $\int_{z}^{\infty} \dfrac{c_s}{H(z)} dz$ decreases. This reduces the length 
scale for baryon drag $r_{\rm drag}$.
As discussed earlier, as BAO measurements constrain the ratio $r_{\rm drag}/D(z)$, for $z 
\simeq z_d$, $r_{\rm drag}$ can be (slightly) reduced in this scenario. As $D \propto 1/H_0$, 
this ensures that BAO observations remain consistent with a large $H_0$. The depletion of 
CDM can be consistent with an enhanced late-time expansion rate shifts the inferred 
Hubble constant $H_0$ toward higher values when fitting cosmological data, offering a 
pathway to reduce the Hubble tension. Finally, the additional DR modifies the late 
Integrated Sachs--Wolfe (ISW) effect and CMB lensing, providing observational constraints 
on the $f {\rm DCDM}$. Further, the enhancement of radiation leads to depletion in the 
growth of density perturbations, especially at small length scales. However, for $\Gamma_{\rm dcdm} 
\simeq H_{EQ}$, this is negligible. \footnote{In contrast, decays at much later times 
primarily suppress clustering without substantially altering $r_{\rm drag}$ \cite{BOSS:2014hhw}.}

Beyond the background evolution, the decay into DR alters the evolution of cosmological 
perturbations through explicit energy--momentum transfer between the DCDM component and DR. 
An analysis of the evolution of the density and velocity perturbations of various species 
is, therefore, essential for reliable estimations of CMB anisotropies and large-scale structure 
observables. We adopt the synchronous-gauge formulation for DCDM, following Ref.\citep{Audren2014}, 
where a single additional parameter $\Gamma_{\rm dcdm}$ was considered. This framework has 
been extended by including the fractional abundance of DCDM ($f_{\rm dcdm}$) \cite{Poulin:2016nat,Xiao:2019ccl}.  
We adopt the synchronous gauge and follow the notations of Ref.~\cite{Ma:1995ey}. The 
perturbation equations governing the evolution of the density contrast $\delta_{\rm dcdm}$ 
and velocity divergence $\theta_{\rm dcdm}$ of the DCDM components are given by, 
\begin{align}
\delta'_{\rm dcdm} &= -\theta_{\rm dcdm} - \frac{1}{2} h',
\label{eq:dcdm-delta} \\
\theta'_{\rm dcdm} &= -\mathcal{H} \theta_{\rm dcdm}.
\label{eq:dcdm-theta}
\end{align}

In the above equation, $h'$ denotes the derivative of the scalar perturbation in the synchronous 
gauge. Note that, in the synchronous gauge the perturbation equations for DCDM are 
unaffected by the decay term. As the CDM perturbations constitute equal-time hypersurface, 
and DCDM and CDM are both at rest in the comoving frame, thus, the decay of DM affects the 
background density, and the density contrast of DCDM in the same way leading to cancellation
in the ratio $\delta_{\rm dcdm}$. The effect of the decay term, which, for example, appears in 
the Newtonian gauge, depletes the density contrast $\delta_{\rm dcdm}$ \cite{Poulin:2016nat}. The Euler 
equation for DCDM remains unchanged, as decay process does not affect the velocity divergencies.

The relativistic decay products contribute as DR. The perturbation equations governing 
the evolution of the lowest-order moments, including the density contrast $\delta_{\rm dr}$, 
velocity divergence $\theta_{\rm dr}$, and shear stress $\sigma_{\rm dr}$, are given by 
\begin{align}
\delta'_{\rm dr} &= - \frac{2}{3} h' + a \Gamma_{\rm dcdm} \frac{\rho_{\rm dcdm}}{\rho_{\rm dr}} 
(\delta_{\rm dcdm}-\delta_{\rm dr}),
\label{eq:dr-delta} \\
\theta'_{\rm dr} &= k^2 \left( \frac{1}{4} \delta_{\rm dr} - \sigma_{\rm dr} \right) - a \Gamma_{\rm dcdm} \frac{\rho_{\rm dcdm}}{\rho_{\rm dr}} \theta_{\rm dcdm},
\label{eq:dr-theta} \\
\sigma'_{\rm dr} &= \frac{4}{15} \theta_{\rm dr} - \frac{3}{5} k F_{{\rm dr},3} - \frac{2}{15} h' - \frac{1}{5} \eta' - a \Gamma_{\rm dcdm} \frac{\rho_{\rm dcdm}}{\rho_{\rm dr}} \sigma_{\rm dr},
\label{eq:dr-sigma}
\end{align}
where $\eta$ is the traceless scalar metric perturbation in the synchronous gauge, and $F_{{\rm dr},3}$ is 
the $\ell=3$ multipole moment of the perturbed DR distribution function. The decay-induced source terms for 
DR signifies the transfer of energy and momentum at the level of perturbation. The higher-order multipoles 
($\ell \geq 3$) evolve according to the standard Boltzmann hierarchy for massless particles. As mentioned, 
at the very early times  $\mathcal{H} \gg \Gamma_{\rm dcdm} {a}$, the probability of decay is 
negligible. Thus, in the CDM sector, the initial conditions are set by the standard adiabatic initial 
conditions of $\Lambda$CDM. In this regime, the perturbations of the DCDM component are indistinguishable 
from those of SCDM, while DR perturbations are initially suppressed. On the contrary, for $\Gamma_{\rm dcdm} 
\gg \dfrac{\mathcal{H}}{a}$, the CDM comprises of SCDM, as most of DCDM would have decayed around  
$\Gamma_{\rm dcdm} \simeq \dfrac{\mathcal{H}}{a}$. In the present context, the priors on the parameter 
$\Gamma_{\rm dcdm}$ ensures that the range of $\Gamma_{\rm dcdm}^{-1}$ includes the radiation--matter 
equality epoch.

\subsection{Dark Energy Parameterization}
\label{subsec:de_param}

Within the standard $\Lambda$CDM framework, the accelerated expansion of the Universe 
\citep{Riess1998} is attributed to a cosmological constant $\Lambda$, characterized by a 
time-independent equation of state $w = -1$. This simplest empirical framework, while consistent 
with current observational data, raises several concerns, most notably the well-known cosmological 
constant problem and the coincidence problem \citep{Weinberg1989,Carroll:2000fy}. These shortcomings 
have motivated extensive investigations into dynamical DE models \cite{Peebles:2002gy}, potentially offering 
a more satisfactory description of the accelerated expansion. 

In this context, a natural and widely adopted approach is to parameterize the DE equation-of-state 
parameter, defined as $w(a) \equiv p_{\rm de}/\rho_{\rm de}$, where $p_{\rm de}$ and $\rho_{\rm de}$ 
are the pressure and energy density of the DE, respectively. Such phenomenological parameterizations 
allow for a model-independent estimation of possible deviations from a cosmological constant 
in a systematic and unbiased manner. In the present context, we adopt the following parameterization,
\begin{equation}
    w(a) = w_0 \,,
\label{eq:w-constant}
\end{equation}
where $w_0$ is a free parameter, and does not vary with time. This corresponds to a one-parameter 
extension in the DE sector, as compared to the $\Lambda$CDM. It is worth mentioning that $w_0 = -1$ 
reproduces the cosmological constant, the DE density scales as $\rho_{\rm de} \propto a^{-3(1+\omega_0)}$.
The regime $w_0 > -1$ describes quintessence-like behavior, in which the DE density decreases 
with time, whereas $w_0 < -1$ corresponds to the phantom regime, associated with a DE density that grows 
with the expansion of the Universe \cite{PhysRevD.37.3406,PhysRevLett.80.1582,PhysRevLett.82.896,Caldwell:1999ew,Peebles:2002gy,Stefancic:2005cs,Copeland:2006wr,Creminelli:2008wc}. The latter scenario violates the null energy condition, 
potentially signaling the need for exotic physics beyond standard scalar field models 
\citep{Caldwell:1999ew}. Despite its simplicity, this one-parameter extension provides a valuable 
first test of deviations from $\Lambda$CDM and serves as a useful baseline. In this article, we 
also comment on a two-parameter extended DE sector,  the Chevallier--Polarski--Linder (CPL) 
parameterization \citep{CPL2001,Linder2003}, which introduces a linear dependence on the scale 
factor $a$,
\begin{equation}
w(a) = w_0 + w_a\,(1 - a) \,,
\label{eq:cpl}
\end{equation}
where $w_0$ is the present-day value of the equation of state and $w_a$ quantifies the rate of its 
dynamical evolution. At the present epoch ($a = 1$), this expression reduces to $w = w_0$, while at 
early times ($a \to 0$) it asymptotes to a finite value $w_0 + w_a$. The respective 
DE density scales as, 
$ \rho_{\rm de}(a) \propto  a^{-3(1+w_0+w_a)}\exp\!\left[-3\,w_a\,(1-a)\right]$. This reduces to the 
$\Lambda$CDM result for $w_0 = -1$ and $w_a = 0$. These DE modifications are implemented at the 
level of the background expansion history and are subsequently combined with the two-parameter 
($f_{\rm dcdm}$, $\Gamma_{\rm dcdm}$) extension of CDM sector described in Sec.~\ref{subsec:dcdm}, 
with the goal of constructing a comprehensive and physically motivated framework for addressing the 
$H_0$ tension from both the early- and late-time sides simultaneously.

\section{Results}
\label{sec:results}

\begin{table*}[htbp]
\centering
\renewcommand{\arraystretch}{1.35}
\caption{Flat prior ranges adopted for the parameters that extend the
         baseline $\Lambda$CDM model. The prior on $\log_{10}(\Gamma_{\rm dcdm})$ 
         is chosen to ensure that decays
         occur around or before radiation--matter equality.}
\label{tab:priors}
\begin{tabular}{lcc}
\hline\hline
Parameter & Symbol & Prior range \\
\hline
Decay rate (log)       & $\log_{10}(\Gamma_{\rm dcdm}/\mathrm{km\,s^{-1}\,Mpc^{-1}})$ & $[5.5,\;8.5]$ \\
Decaying DM fraction   & $f_{\rm dcdm}$     & $[0,\;0.2]$   \\
DE EoS        & $w_0$              & $[-3.0,\;1.0]$ \\
DE EoS slope  & $w_a$              & $[-3.0,\;2.0]$ \\
\hline\hline
\end{tabular}
\end{table*}

\begin{table}[htbp]
\centering
\renewcommand{\arraystretch}{1.35}
\caption{Cosmological datasets used in this analysis.}
\label{tab:datasets}
\begin{tabular}{llll}
\hline\hline
Sl. no. & Dataset & Likelihood component & Multipole range / Redshift \\
\hline
1   &\texttt{Planck 2018} \citep{Planck2018} 
    & \texttt{Low-$\ell$ TT}           & $\ell = 2$--$29$          \\
    &  &\texttt{Low-$\ell$ EE (Sroll2)}  & $\ell = 2$--$29$          \\
    &  &\texttt{High-$\ell$ TT (ACT cut)}& $\ell \leq 1000$          \\
    &  &\texttt{High-$\ell$ TE/EE}       & $\ell \leq 600$           \\
2   &\texttt{ACT DR6} \citep{AtacamaCosmologyTelescope:2025blo}
    & \texttt{High-$\ell$ TT/TE/EE}    & $\ell \gtrsim 600$        \\
3   & \texttt{CMB lensing}
    & \texttt{CMB} lensing reconstruction & $\ell \leq 4000$         \\
4   & \texttt{DESI DR2} \citep{DESI:2025fii}
    & \texttt{BAO} distance measurements  & All redshift bins     \\
5   & \texttt{SH0ES + Pantheon+} \citep{Brout:2022vxf}
    & Cepheid distance ladder  & $0 < z \lesssim 2.3$ \\
    & & + SN Ia magnitude &  \\
\hline\hline
\end{tabular}
\end{table}

In this section, we present the observational constraints on the extensions of the 
standard model of cosmology, as described in Sec.~\ref{sec:model} in light of 
cosmological data. In particular, we  discuss the implications on the CMB anisotropies, 
the matter power spectrum, and on the cosmological tensions. Broadly, we analyze the 
following scenarios:
\begin{enumerate}
    \item \texttt{Model-A} : Fractional DCDM ($f {\rm DCDM}$) (with $w_0=-1$) \\
          A fraction $f_{\rm dcdm}$ of cold dark matter decays into invisible DR with a constant 
          decay width $\Gamma_{\rm dcdm}$, with no free parameter in the DE sector (i.e., cosmological 
          constant).

    \item \texttt{Model-B} : Fractional DCDM ($f {\rm DCDM}$) + evolving DE \\
          The same decaying dark matter component is combined with a constant equation of state 
          $w_0$ for DE. Also, we briefly touch upon the implications for the CPL parametrization 
          $(w_0,\,w_a)$.
\end{enumerate}

Parameter inference is performed through a Bayesian Markov Chain Monte Carlo (MCMC) analysis 
using the publicly available package \texttt{COBAYA} \citep{Torrado:2020dgo} interfaced with 
the publicly available Boltzmann equation solver \texttt{CLASS} \cite{blas2011cosmic} where 
the covariant DCDM perturbation equations were implemented. The code was modified extending the 
analysis to the fractional-decaying DM, as described in Ref.~\cite{Audren2014}. We vary the six 
baseline $\Lambda$CDM parameters $\{\Omega_b h^2,\,\Omega_{\rm cdm} h^2,\,100\theta_{\rm MC},\,\tau_{\rm
reio},\,\log(10^{10}A_s),\,n_s\}$ together with the model-specific parameters $f_{\rm dcdm}$,
$\log_{10}(\Gamma_{\rm dcdm}/\mathrm{km\,s^{-1}\,Mpc^{-1}})$ and $w_0$. Further, $w_a$ is also 
varied while considering CPL parameterization for DE sector. The parameters considered, and the 
respective flat priors have been mentioned in Table~\ref{tab:priors}. The combination of datasets used   
is detailed in Table~\ref{tab:datasets}.  For 
early times, datasets 1-4 have been used. While for late times, dataset 5 
has been added. In this set, \texttt{Pantheon+} supernovae data have been used with Cepheid calibration from \texttt{SH0ES}. We have used \texttt{SH0ES} data
in this context, as it provides robust calibration in the entire low redshift range. Further, $H_0$ obtained by \texttt{SH0ES} and \texttt{H0DN} are very similar \cite{H0DN:2025lyy,Riess2022}. 
To assess the convergence of the chains, the 
Gelman--Rubin criterion $R-1<0.01$ \cite{Gelman92,Trotta:2008qt} has been followed. For model comparison, the 
difference in best-fit $\chi^2$, $\Delta\chi^2 \equiv \chi^2_{\rm model}-\chi^2_{\Lambda\rm CDM}$, 
has been used, where negative values favour the extended model.

\subsection{Implications on the Parameters in the Dark Sector}
\label{subsec:dcdm_w}

\begin{figure}[t]
    \centering
    \includegraphics[width=\columnwidth]{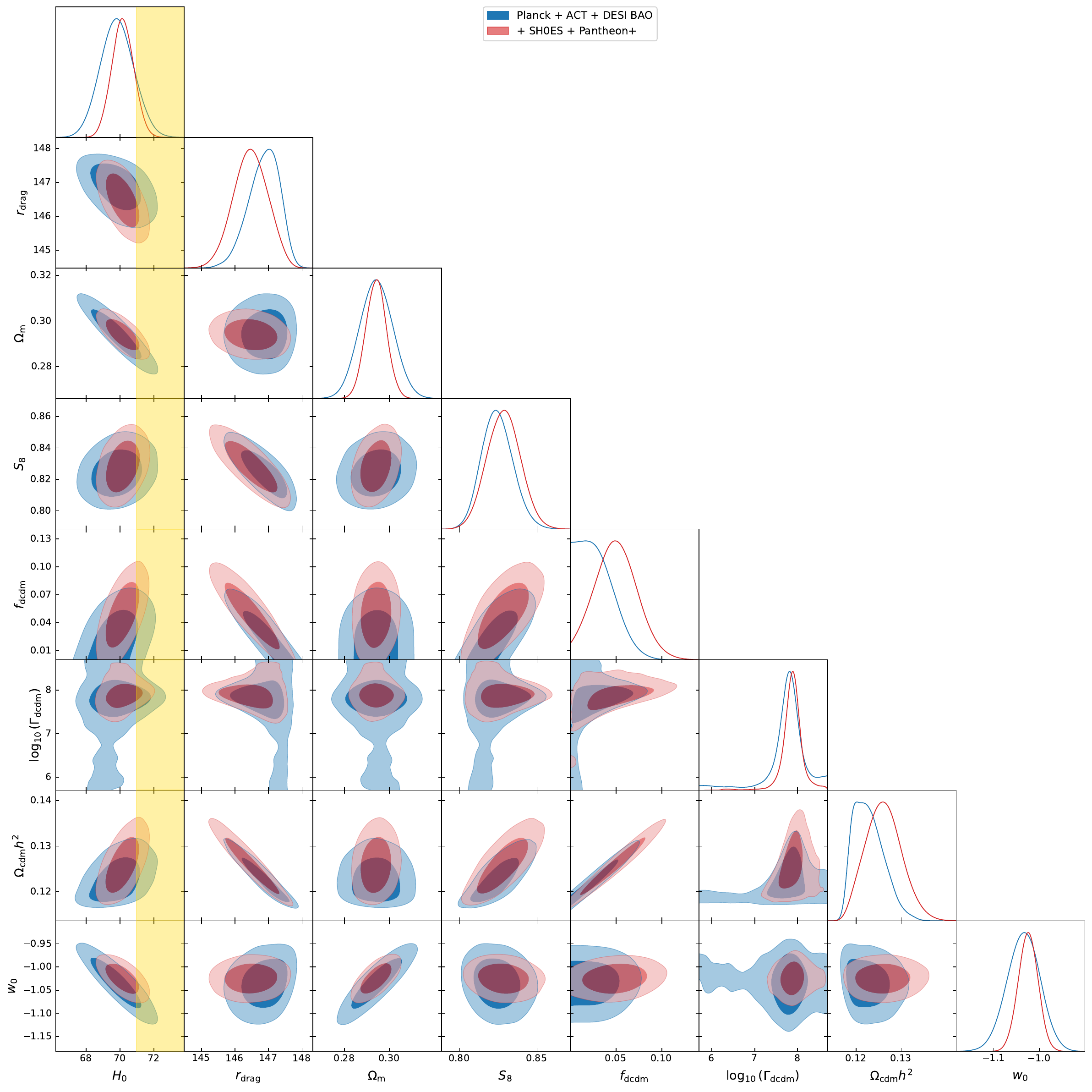}
    \caption{Triangle plot for \texttt{Model-B} showing the marginalized posterior
distributions for the parameters $\omega_{\rm cdm}$, $\Omega_{\rm m}$,
$\log\Gamma_{\rm dcdm}$ (log decay rate), $S_8$, $H_0$, $r_{\rm drag}$
(sound horizon at drag epoch), $f_{\rm dcdm}$ (fraction of decaying dark matter),
and $w_0$ (dark energy equation-of-state parameter).
The \textbf{blue} contours correspond to the dataset combination
\texttt{Planck +ACT DR(6)+DESI BAO (DR2)}, while the \textbf{red} contours correspond to
\texttt{Planck + ACT (DR6)+DESI BAO (DR2) +SH0ES+Pantheon\texttt{+}}.
All panels display both the $1\sigma$ ($68\%$,C.L.) and
$2\sigma$ ($95\%$,C.L.) confidence regions.
The \textbf{gold shaded band} represents the $2\sigma$ SH0ES constraint on $H_0$.
}
    \label{fig:triangle_dcdm_w0}
\end{figure}

\begin{table*}[htbp]
\centering
\footnotesize
\setlength{\tabcolsep}{3pt}
\caption{Mean values and $1\sigma$ uncertainties of cosmological parameters obtained from the combined \texttt{Planck 2018 + ACT DR6 + DESI DR2 (BAO)} data. The quantity $\Delta\chi^2 \equiv \chi^2_{\rm model} - \chi^2_{\Lambda\rm CDM}$ is computed with respect to the baseline $\Lambda$CDM model. Upper limits on $f_{\rm dcdm}$ and $\Sigma m_\nu$ are at 95\% CL.}
\label{tab:all_models}
\resizebox{\textwidth}{!}{%
\begin{tabular}{lcccc}
\hline\hline
Parameter
& $\Lambda$CDM
& \texttt{Model-A}
&  \texttt{Model-B}
& \texttt{Model-B}+$\sum m_\nu$ \\
\hline
 $A_s\times10^{9}$
& $2.10\pm0.03$
& $2.147\pm0.026$
& $2.135\pm0.026$
& $2.136\pm0.025$ \\
$n_s$
& $0.974\pm0.003$
& $0.9797\pm0.0048$
& $0.9775\pm0.0049$
& $0.9781\pm0.0046$\\
$100\,\theta_s$
& $1.0411\pm0.0003$
& $1.0419\pm0.0003$
& $1.0419\pm0.0003$
& $1.04186\pm0.00026$ \\
$\Omega_b h^2$
& $0.0224\pm0.00015$
& $0.02251\pm0.00012$
& $0.02249\pm0.00012$
& $0.02250\pm0.00012$\\
$\Omega_{\rm cdm}h^2$
& $0.120\pm0.001$
& $0.1217\pm0.0034$
& $0.1222\pm0.0033$
& $0.1226\pm0.0032$\\
$\tau_{\rm reio}$
& $0.0624\pm0.007$
& $0.0635\pm0.0064$
& $0.0615\pm0.0069$
& $0.0613\pm0.0062$ \\
$w_0$
& $-1$ (fixed)
& $-1$ (fixed)
& $-1.035\pm0.036$
& $-1.025\pm0.038$\\
$f_{\rm dcdm}$
& -- & $<0.064$  & $<0.062$
& $<0.061$\\
$\log_{10}\!\left(\dfrac{\Gamma_{\rm dcdm}}{\rm km\,s^{-1}\,Mpc^{-1}}\right)$
& -- & $7.72\pm0.52$  & $7.75\pm0.50$
& $7.72\pm0.54$ \\
$\sum m_\nu \,[{\rm eV}]$
&0.06 (fixed) & 0.06 (fixed) &  0.06 (fixed)
& $<0.10$ \\
\hline
$\mathbf{H_0}$
& $\mathbf{68.38\pm0.28}$
& $\mathbf{68.98\pm0.47}$
& $\mathbf{69.83\pm0.98}$
& $\mathbf{69.70\pm0.99}$ \\
$S_8$
& $0.817\pm0.007$
& $0.823\pm0.014$
& $0.824\pm0.014$
& $0.834\pm0.013$ \\
$\Omega_\Lambda (\Omega_{fld})$
& $0.690\pm0.04$
& $0.6951\pm0.0053$
& $0.7059\pm0.0073$
& $0.7053\pm0.0074$ \\
$r_{\rm drag}$ [Mpc]
& $147.58\pm0.20$
& $146.98\pm0.45$
& $146.88\pm0.45$
& $146.88\pm0.44$ \\
\hline
$\Delta\chi^2$
& $0$
& $-2.90$
& $-3.60$
& $-5.2$ \\
\hline\hline
\end{tabular}%
}
\end{table*}

\begin{table*}[htbp]
\centering
\footnotesize
\setlength{\tabcolsep}{5pt}
\caption{Mean values and $1\sigma$ uncertainties obtained from the combined \texttt{Planck+ACT+DESI\,DR2+SH0ES+Pantheon+} dataset.}
\label{tab:shoes_pantheon}
\resizebox{0.8\textwidth}{!}{%
\begin{tabular}{lcc}
\hline\hline
 & \multicolumn{2}{c}{\texttt{Planck+ACT+DESI\,DR2+SH0ES+Pantheon+}} \\
\cmidrule(lr){2-3}
Parameter
 & $\Lambda$CDM
 & \texttt{Model-B} \\
\hline
$\log(10^{10} A_\mathrm{s})$        
& $3.051 \pm 0.012$          
& $3.050 \pm 0.011$ \\
$n_\mathrm{s}$                       
& $0.9762 \pm 0.0031$        
& $0.9875 \pm 0.0050$ \\
$100\,\theta_\mathrm{s}$             
& $1.04187 \pm 0.00024$      
& $1.04213 \pm 0.00025$ \\
$\Omega_\mathrm{b} h^2$              
& $0.022617 \pm 0.000107$    
& $0.022549 \pm 0.000113$ \\
$\Omega_\mathrm{cdm} h^2$              
& $0.11690 \pm 0.00062$      
& $0.13019 \pm 0.00317$ \\
$\tau_\mathrm{reio}$                 
& $0.0659 \pm 0.0064$        
& $0.0623 \pm 0.0059$ \\
$w_0$                                
& $-1$ (fixed)         
& $-1.025 \pm 0.021$ \\
$w_a$                                
& ---                        
& --- \\
$f_{\rm dcdm}$                    
& --                  
& $0.073 \pm 0.017$ \\
$\log_{10}(\Gamma_{\rm dcdm}/{\rm km\,s^{-1}\,Mpc^{-1}})$
& --            
& $7.85 \pm 0.25$ \\
$\mathbf{H_0}\;[\mathbf{km\,s^{-1}\,Mpc^{-1}}]$
& $\mathbf{68.60 \pm 0.26}$  
& $\mathbf{70.20 \pm 0.66}$ \\
$S_8$                                
& $0.807 \pm 0.007$          
& $0.828 \pm 0.010$ \\
$r_\mathrm{drag}\;[\mathrm{Mpc}]$   
& $147.66 \pm 0.19$          
& $146.45 \pm 0.50$ \\
\hline
$\Delta \chi^2$                      
& $0$                        
& $-8.8$ \\
\hline\hline
\end{tabular}%
}
\end{table*}
In the following, we discuss the extended Dark Sector models discussed 
above in light of the cosmological data, as mentioned in Table \ref{tab:datasets}.

\paragraph{\texttt{Model-A} [$f {\rm DCDM}$ benchmark (with $w_0=-1$)]:}
We first consider the simplest extension of $\Lambda$CDM, among the models considered 
in this study, i.e., with $f$DCDM. As mentioned earlier, it is referred to as \texttt{Model-A}. 
The equation of state for DE is fixed at $w_0  = -1$. For the analysis, data sets 1 to 4, as 
mentioned in Table \ref{tab:datasets} have been used. The best fit values, and the allowed 
range of the parameters have been mentioned in the second column 
of Table~\ref{tab:all_models}. 

The best fit value for $\Gamma_{\rm dcdm}$ is given by $\Gamma_{\rm dcdm} = 5.25\times10^{7}\; 
\mathrm{km\, s^{-1}\, Mpc^{-1}}$, and the parameter lies approximately between $1.5\times10^{7}- 1.8\times 10^{8}\; \mathrm{km\, s^{-1}\, Mpc^{-1}}$ at 68\% c.l. while the decaying fraction is constrained as 
$f_{\rm dcdm} < 0.064$ (at 95\% c.l.). The Hubble parameter today ($H_0$) shifts upward 
to $H_0 = 68.98 \pm 0.47\; \mathrm{km\,s^{-1}\,Mpc^{-1}}$, reducing the discrepancy with 
the \texttt{H0DN} measurement \citep{H0DN:2025lyy} from approximately $6.0\sigma$ to 
$4.8\sigma$. The best fit value of $\Gamma_{\rm dcdm}$ corresponds to a lifetime of the decaying 
component of approximately $\mathcal{O}(10^4)$ years, which is well after the Big Bang 
Nucleosynthesis (BBN) epoch, and around the radiation--matter equality epoch. The sound 
horizon at baryon drag is mildly compressed to $r_{\rm drag} = 146.98 \pm 0.45\;\mathrm{Mpc}$, 
approximately $0.6\;\mathrm{Mpc}$ below the $\Lambda$CDM value, in accordance with the 
discussion in Sec.~\ref{sec:model}. 

In the following, we comment on the BBN constraints in this context. The success of BBN 
imposes constraints on the Hubble parameter, and consequently, the number of relativistic 
species (with the same temperature as the SM plasma)  \cite{mangano2011robust,Cyburt:2015mya}. However, for $f_{\rm dcdm} 
\simeq 0.06$, around $z \simeq 10^9$ (temperature 
$T \simeq \mathcal{O}(1-10)$ MeV) the relative change in the Hubble parameter during BBN is given by
$\left.\frac{H^{f \rm DCDM} - H^{\rm \Lambda CDM}}{H^{\rm \Lambda CDM}} 
\right|_{(T\simeq 1 {\rm MeV})} \ll 0.001$. This is because, during BBN DCDM behaves as matter, 
and as it contributes less than 10\% of the CDM, its contribution to the expansion rate at 
$T\simeq 1 ~{\rm MeV}$, which falls in the radiation-dominated epoch, remains insignificant. 
For the same reason, there is no contribution to the effective number of additional 
relativistic species $\Delta N_{\rm eff}$ at the onset of BBN in this scenario. It is also 
worth noting that, as the decay width of the DCDM component is very small, the respective 
interaction strength of DCDM and DR is negligibly small, and thus, in the early Universe 
DCDM and DR components would not be in thermal equilibrium in this scenario. This is in contrast to the models described in Refs.~\cite{Cyr-Racine:2013fsa,Garny:2025kqj,Garny:2025szk}, where CDM–DR interactions are appreciable in the early Universe. However, the stability of CDM is typically ensured by underlying symmetry principles, and no decaying component is present. Thus,  
appreciable production of the DR component does not happen before the BBN epoch in this 
scenario. However, as the DCDM decays to DR around the radiation-matter equality epoch, 
it produces additional radiation. Consequently, the energy density in the radiation bath 
increases. As the decay process happens towards the end of the radiation-dominated epoch, 
$f_{\rm dcdm} \lesssim 10\%$ implies that the contribution from the DCDM to the total 
energy density is less than $5\%$ approximately. Consequently, at the epoch of recombination, 
which is in the matter-dominated epoch, the respective relativistic decay products contribute 
less than $1.5\%$, as the radiation energy density depletes by a factor of $a^{-1}$ faster 
compared to the matter density. Therefore, the constraint on $\Delta N_{\rm eff}$ ($<0.15$ at recombination) from 
ACT is respected for the 95\% upper limit on $f_{\rm dcdm}$, as depicted in 
Table \ref{tab:all_models}. The sum of the (three species of) neutrino masses 
$\Sigma m_{\nu}$ is set to 0.06 eV for 
the present analysis.

\paragraph{\texttt{Model-B} [$w_0 +f {\rm DCDM}$ model]:}
Introducing a constant DE equation of state alongside the $f {\rm DCDM}$ component yields 
a more pronounced improvement, as far as the Hubble tension is concerned. As depicted in 
Table \ref{tab:all_models}, with datasets 1-4 (see Table \ref{tab:datasets}), there is
a mild but statistically non-zero preference for a decaying fraction $f_{\rm dcdm}$,
\begin{equation}
    f_{\rm dcdm} = 0.027 \pm 0.019
    \quad \bigl(< 0.062\ \text{at}\ 95\%\ \mathrm{C.L.}\bigr).
\label{eq:fdcdm_w0}
\end{equation}
The best-fit value for the decay width $\Gamma_{\rm dcdm}$ such that the life-time $\Gamma_{\rm dcdm}^{-1}$ 
of the DCDM particle is around the radiation--matter equality epoch, and is given by, 
\begin{equation}
    \log_{10}\!\left(
        \frac{\Gamma_{\rm dcdm}}{\mathrm{km\,s^{-1}\,Mpc^{-1}}}
    \right) = 7.75 \pm 0.50 \,.
\label{eq:Gamma_w0}
\end{equation}
In this scenario, for the  $r_{\rm drag}$ is reduced by a further ${\sim}0.1\;\mathrm{Mpc}$ 
relative to the best fit point $\texttt{Model-A}$. The  Hubble parameter in the present epoch 
is increased to, $H_0 = 69.83 \pm 0.98\;\mathrm{km\,s^{-1}\,Mpc^{-1}} \,.$

This shows a significant upward shift as compared to its value in $\Lambda$CDM, as well 
from \texttt{Model-A}, reducing the residual tension with the \texttt{H0DN} local measurement 
\cite{H0DN:2025lyy} to ${\sim}2.9\sigma$. The inferred equation of state of DE is given by 
$w_0 = -1.035 \pm 0.036$, is consistent with a cosmological constant at the $1 \sigma$ level. 
It displays a marginal phantom preference that provides additional late-time expansion 
without any significant effect on the CMB acoustic peaks. The matter clustering amplitude 
remains essentially unaffected, $S_8 = 0.829 \pm 0.014$, which is consistent with large-scale 
structure observations. The overall goodness of fit improved compared to $\Lambda$CDM 
($\Delta\chi^2 \approx -3.6$). This suggests that the upward shift in
$H_0$ arises from early- and late-time effects, providing a better fit to both 
CMBR and BAO. The value of $H_0$ obtained in this framework 
is in good agreement with the \texttt{SH0ES} measurement and is within less than 2.5$\sigma$, significantly reducing the tension of $H_0$. Consequently, 
in addition to the datasets 1-4 in Table \ref{tab:datasets}, we considered 
the \texttt{SH0ES + 
Pantheon+}, as mentioned in the fifth row of the same table. This further enhances the Hubble parameter 
in the present epoch $H_0 = 70.20 \pm 0.66;\mathrm{km\,s^{-1}\,Mpc^{-1}} \,$, raising the central value and reducing the uncertainty.The parameter $r_{\rm drag}$ is decreased to 146.45 Mpc.
All the relevant parameters have been shown in Table \ref{tab:chi2_combined}. This improvement 
is primarily due to the preference of the late-time datasets to a higher expansion rate. The 
goodness of fit improved compared to $\Lambda$CDM ($\Delta\chi^2 \approx -8.8$).
It is worth noting that, as shown in Figure.\ref{fig:triangle_dcdm_w0}, the posteriors of the decay 
width $\Gamma_{\rm dcdm}$ show clear peaks for both \texttt{Model-A} and \texttt{Model-B}, demonstrating a 
preference for the decaying component around the epoch of radiation-matter equality epoch. 
The 2-$\sigma$ contours include the entire prior range, while considering datasets sensitive to the early universe i.e., CMB (including lensing) and BAO (serial no. 1–4 in Table.~\ref{tab:datasets}). This is because the prior range in $f_{\rm dcdm}$ 
includes zero, and $\Gamma_{\rm dcdm}$ becomes unconstrained as $f_{\rm dcdm} \rightarrow 0$. The contours for $f_{\rm dcdm}$ change substantially after including the datasets  SH0ES+PantheonPlus (serial no. 5 in Table.\ref{tab:datasets})  which include the measurement of $H_0$ via Cepheid-calibrated Supernova-type IA  and are sensitive to the late Universe. In this case, $f_{\rm dcdm} \rightarrow 0$ becomes disfavored at 1-$\sigma$, and consequently, the contours for $\Gamma_{\rm dcdm}$ is well constrained around the best-fit 
values. Note that, for a given $f_{\rm dcdm}$ a very large $\Gamma_{\rm dcdm}$ corresponds to a smaller 
lifetime of DCDM, and thus, it decays well within the radiation-dominated epoch, producing DR. 
Consequently, the effect $H(z)$ and, thus, on $r_{\rm drag}$ is not substantial. For much 
smaller $\Gamma_{\rm dcdm}$, the decay takes place around the recombination epoch, which is disfavored by 
CMBR. Note that, as expected, $f_{\rm dcdm}$ shows a 
positive correlation with $H_0$, and a negative correlation 
with $r_{\rm drag}$. Note that,  as mentioned earlier, including \texttt{SH0ES+ Pantheon+}(mentioned in sl.no 5 in Table.\ref{tab:datasets}); substantially raises $H_0$ and one obtains $H_0=70.20$. The details have been mentioned in Table.~\ref{tab:shoes_pantheon}.

Next, we comment on the 
implications on the sum of the neutrino masses $\Sigma m_{\nu}$ in this context. For our analysis, as mentioned in the third column of Table \ref{tab:all_models}, $\Sigma m_{\nu}$ is 
set to 0.06 eV. As this parameter is correlated with $w_0$~\cite{Elbers:2025vlz,Shao:2024mag}, therefore, an extension of \texttt{Model-B} varying $\Sigma m_{\nu}$ has also been considered.
The results have been stated in the fourth column of Table \ref{tab:all_models}. The upper limit on 
$\Sigma m_{\nu}$ is relaxed to 0.1 eV. 
\footnote{In appendix A, Table \ref{tab:additional_runs}, a one parameter extension of $\Lambda$CDM with 
$\Sigma m_{\nu}$, and a two parameter extension with $w_0, ~ \Sigma m_{\nu}$ 
have been depicted for a reference. The 
relaxation of the (68\%) upper limit of 
$\Sigma m_{\nu}$ is maximal for \texttt{Model-B}.} There is an increase in the decay width $\Gamma_{\rm dcdm}$ 
by about a factor of 1.5, as compared to 
the scenario, where $\Sigma m_{\nu}$ was fixed at 0.06 eV. No significant change in 
the upper limit of $f_{\rm dcdm}$ is noticed. Note that an early decay of the 
DCDM component would correspond to further depletion of the DR, as compared to matter density around the recombination epoch. It (partially) 
contributes to a proportionate increase in the upper limit of $\Sigma m_{\nu}$, enhancing the contribution of the (semi-)relativistic neutrino species around the same epoch. Note that this increase in the upper limit to 0.1eV (at 95\% c.l.) allows the inverted hierarchy in the neutrino sector to remain a viable possibility.

To evaluate the statistical viability of the models beyond goodness-of-fit metrics, we perform a Bayesian model comparison utilizing the \texttt{PolyChord} nested sampling algorithm \citep{Handley:2015fda}.
The computed log-Bayes factors yield strictly $|\ln B| < 1$ (+0.48), denoting an inconclusive statistical preference on the empirical Jeffreys scale. Furthermore, while the incorporation of late-time observational datasets precipitates a marginal posterior shift favouring the extended \texttt{Model-B} cosmology (-0.9), this enhanced phenomenological fit only partially mitigates the inherent Occam penalty exacted by the introduction of additional model parameters.\footnote{The detailed $\chi^2$ contributions for the respective datasets, along with the results of the Bayesian analysis, are presented in Table \ref{tab:chi2_combined} and will be discussed subsequently.} 

Finally, we briefly comment on the extension of \texttt{Model-B}, with evolving DE 
equation of state described by the Chevallier--Polarski--Linder (CPL) parametrization as 
$w(z)= w_0+w_a(1-a)$. Using the datasets 1-4 in Table \ref{tab:datasets}, we obtain 
$H_0 = 65.4 \pm 2.52\;\mathrm{km\,s^{-1}\,Mpc^{-1}}$. There is substantial broadening of the 
$H_0$ posterior as compared to \texttt{Model-A} which likely reflects a degeneracy between $w_a$ and the early-time decay parameters. We will not explore this possibility further in this article. \subsection{\texorpdfstring{Status of $H_0$ (and $S_8$) Tension in the Extended Models}
{Status of H0 (and S8) Tension in the Extended Models}}

As the focus of the present study is to address the cosmological tensions, in the following, we present a discussion on the status of the tensions in the extended 
dark sector frameworks.

\begin{table*}[htbp]
\caption{Constraints on the Hubble constant $H_0$ for \texttt{Model-B} from the combination of \texttt{Planck+ACT (DR6)+DESI(BAO) DR2} data in various cosmological models, along with the residual tension with respect to the local measurement from the \texttt{H0DN} Collaboration \cite{H0DN:2025lyy}, $H_0 = 73.50 \pm 0.81\;\mathrm{km\,s^{-1}\,Mpc^{-1}}$, and the \texttt{SH0ES} Collaboration \cite{Riess2022}, $H_0 = 73.04 \pm 1.04\;\mathrm{km\,s^{-1}\,Mpc^{-1}}$. Tensions are quoted in units of $\sigma$ significance.}
\label{tab:h0_tension}

\centering 
\renewcommand{\arraystretch}{1.3}

\begin{tabular}{lccc}
\hline\hline
Model & $H_0\;[\mathrm{km\,s^{-1}\,Mpc^{-1}}]$ & \texttt{H0DN} tension & \texttt{SH0ES} tension \\
\hline
$\Lambda$CDM & $68.38 \pm 0.28$ & ${\sim}6.0\,\sigma$ & ${\sim}4.3\,\sigma$ \\
\texttt{Model-A} & $68.98 \pm 0.47$ & ${\sim}4.8\,\sigma$ & ${\sim}3.4\,\sigma$ \\
$w_0$CDM & $69.11 \pm 1.00$ & ${\sim}3.4\,\sigma$ & ${\sim}2.8\,\sigma$ \\
\texttt{Model-B} & $69.83 \pm 0.98$ & ${\sim}2.9\,\sigma$ & ${\sim}2.2\,\sigma$ \\
\hline\hline
\end{tabular}
\end{table*}

\begin{figure}[t]
    \includegraphics[width=\columnwidth]{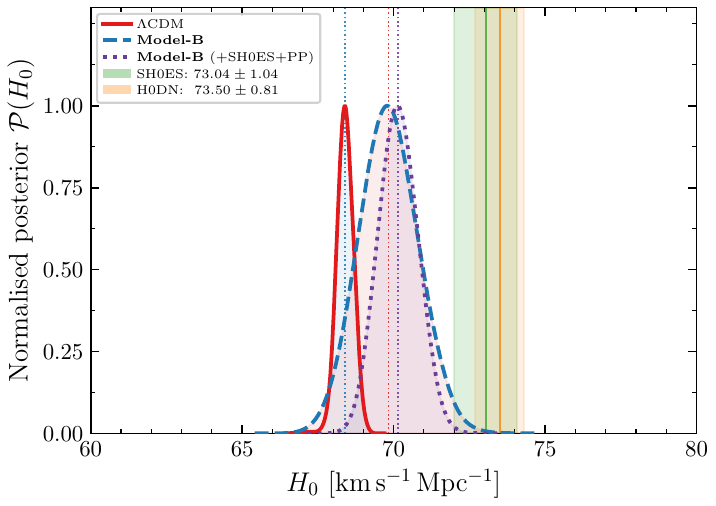}
    \caption{Marginalized posterior distributions of $H_0$ for the $\Lambda$CDM baseline (solid red), \textbf{Model-B} (dashed blue), and \textbf{Model-B + SH0ES (SH0ES+PP)} (dotted purple), where \textbf{PP} refers to \texttt{Pantheon+}. The $\Lambda$CDM and Model-B constraints are obtained using \texttt{Planck\,+\,ACT DR6\,+\,DESI DR2}, while the purple curve additionally includes the SH0ES with Pantheon+. Shaded vertical bands indicate the local distance-ladder measurements 
    from \texttt{SH0ES} $73.04\pm1.04\ \mathrm{km\,s^{-1}\,Mpc^{-1}}$ \cite{Riess2022} and \texttt{H0DN} $73.50\pm0.81\ \mathrm{km\,s^{-1}\,Mpc^{-1}}$ \cite{H0DN:2025lyy}. 
    Thin dotted vertical lines mark the posterior mean values for each model.}
    \label{fig:H0_posterior}
\end{figure}

\begin{figure}[t]
    \centering
    \begin{subfigure}[t]{0.45\linewidth}
        \centering
        \includegraphics[width=\linewidth]{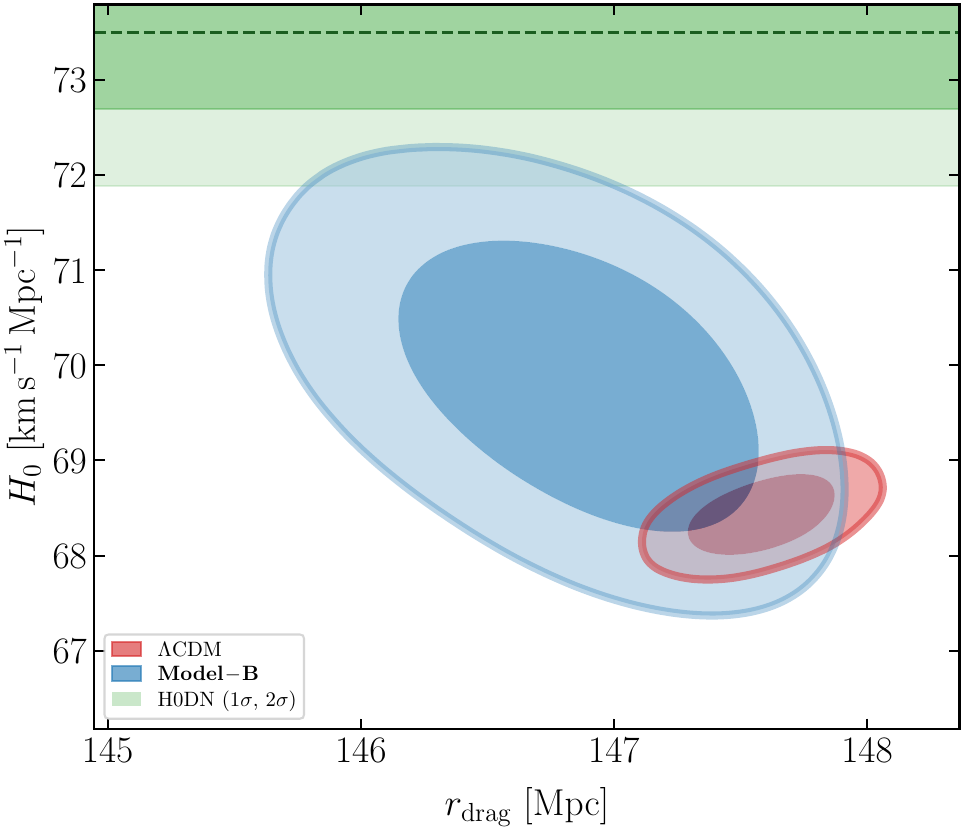}
        \caption{$H_0$ vs $r_{\rm drag}$}
        \label{fig:h0_drag}
    \end{subfigure}
    \hfill
    \begin{subfigure}[t]{0.45\linewidth}
        \centering
        \includegraphics[width=\linewidth]{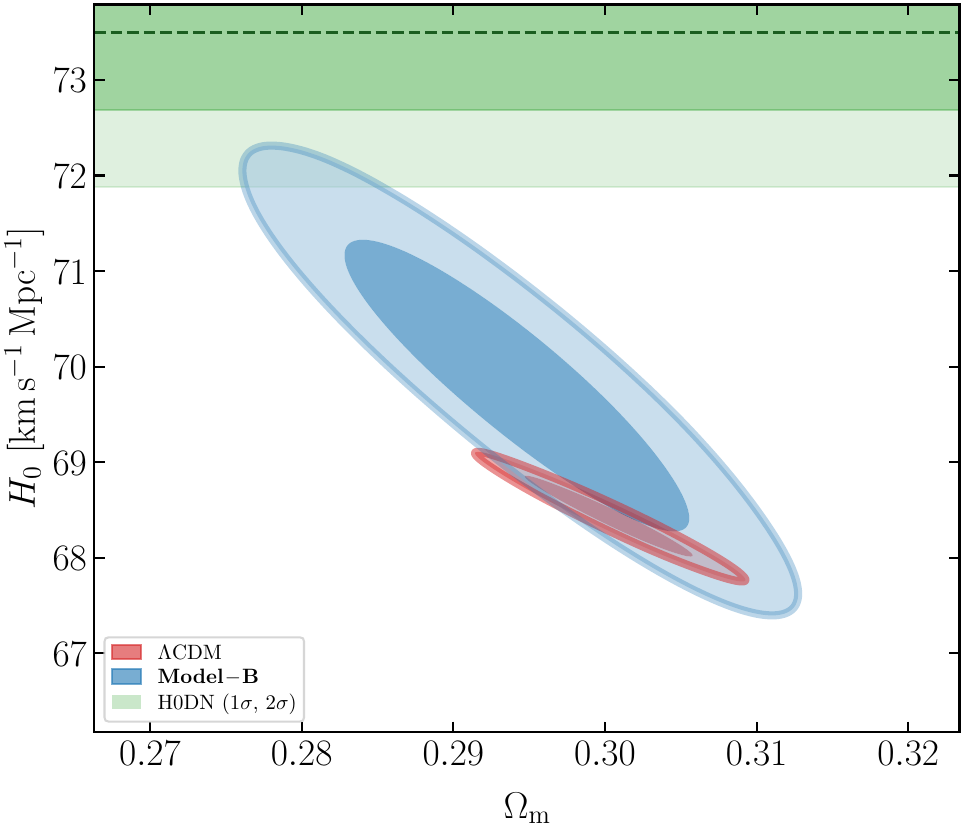}
        \caption{$H_0$ vs $\Omega_m$}
        \label{fig:h0_omega_m_panel}
    \end{subfigure}
    \caption{2D marginalized contours (68\% and 95\%~C.L.) in $\Lambda$CDM (red) and \texttt{Model-B} (blue) with a \texttt{H0DN} ($1-2 \sigma$) on $H_0$ \citep{H0DN:2025lyy}.
    \emph{Left:} $H_0$ vs sound horizon at baryon drag $r_{\rm drag}$.\emph{Right:} $H_0$ vs present-day matter density parameter $\Omega_m$.}
    \label{fig:h0_omega_m}
\end{figure}

Figure~\ref{fig:H0_posterior} shows the marginalized posterior distributions of $H_0$ 
for the $\Lambda$CDM baseline and the $w_0+f {\rm DCDM}$ hybrid model, compared against the 
local distance-ladder measurements from \texttt{SH0ES} \citep{Riess2022} and \texttt{H0DN} \citep{H0DN:2025lyy}. 
The $\Lambda$CDM fit to \texttt{Planck+ACT (DR6)+DESI(BAO) DR2} data, i.e., serial no. 1-4 in Table \ref{tab:datasets}, yields
\begin{equation}   H_0^{\Lambda\mathrm{CDM}} = 68.38 \pm 0.28 \ \mathrm{km\,s^{-1}\,Mpc^{-1}} \quad (68\%\ \mathrm{C.L.}),
\end{equation}
in tension with the local distance-ladder measurements at the $\sim 6.0\sigma$ level. In \texttt{Model-B}, the $H_0$ posterior shifts towards higher values,
\begin{equation}
    H_0^{\rm Model-B} = 69.83 \pm 0.98 \ \mathrm{km\,s^{-1}\,Mpc^{-1}} \quad (68\%\ \mathrm{C.L.}),
\end{equation}
reducing the tension with the local measurements to $\sim 2.9\sigma$, can be seen in Figure~\ref{fig:H0_posterior}. This 
shift is driven by the interplay between the DCDM component, parametrized by the 
decay rate $\Gamma_{\rm dcdm}$ and DCDM fraction $f_{\rm dcdm}$, and the 
DE equation-of-state parameter $w_0$, which together modify the late-time 
expansion history in a manner that partially alleviates the $H_0$ discrepancy. 

Table~\ref{tab:h0_tension} summarises the Hubble constant posterior
for all the models considered in this work, using the combined dataset
\texttt{Planck+ACT (DR6)+DESI(BAO) DR2}, i.e., serial no. 1-4, as mentioned in Table \ref{tab:datasets}. The successive extensions of the
dark sector shift the inferred values progressively upward: introducing
\texttt{Model-A} raises $H_0$ to $68.98 \pm
0.47\;\mathrm{km\,s^{-1}\,Mpc^{-1}}$, reducing the tension to
${\sim}4.8\,\sigma$; allowing a constant DE equation of state
$w_0$ without $f {\rm DCDM}$ gives $H_0 = 69.11 \pm
1.00\;\mathrm{km\,s^{-1}\,Mpc^{-1}}$ (${\sim}3.4\,\sigma$); and the
combination of both --- the \texttt{Model-B} --- yields the largest shift,
$H_0 = 69.83 \pm 0.98\;\mathrm{km\,s^{-1}\,Mpc^{-1}}$, lowering the
residual tension to ${\sim}2.9\,\sigma$. With the addition of \texttt{SH0ES} and \texttt{Pantheon+} data, as shown in Table~\ref{tab:shoes_pantheon}),    \texttt{Model-B} yields $H_0 = 70.20 \pm 0.66\;\mathrm{km\,s^{-1}\,Mpc^{-1}}$, significantly closer to local distance-ladder measurements.

The progressive improvement in $H_0$ follows directly from the physical 
mechanisms described in Sec.~\ref{sec:model}: DM decays near radiation--matter equality compress $r_{\rm drag}$, while DE amplifies the late-time expansion. This is illustrated in Figure~\ref{fig:h0_omega_m}. The left panel shows a negative correlation between $H_0$ and $r_{\rm drag}$ in \texttt{Model-B}, and a bigger parameter region favoured by the data. The right panel demonstrates $H_0$ vs $\Omega_m$ in \texttt{Model-B}, where a larger region is favoured by the data 

Next, we comment on the $S_8$ tension in the context of the extended models considered in this 
work. The $S_8$ values inferred across all models remain in mild tension with weak-lensing 
measurements from surveys such as \texttt{KiDS} and \texttt{DES}, which typically report 
$S_8 \sim 0.76$--$0.78$~\cite{DES:2026fyc}, while results from the KiDS Legacy Survey indicate that the previously reported $S_8$ tension is no longer significant \citep{Stolzner:2025htz}. The $f {\rm DCDM}$ models yield slightly higher 
$S_8 \approx 0.828$--$0.829$ compared to $\Lambda$CDM ($S_8 = 0.817$). 
The introduction of DCDM alone does not fully alleviate this tension. 
However, interactions between the residual SCDM and baryons 
with an effective cross-section ($\sigma_{\rm dmeff}$) can suppress small-scale power 
sufficiently to ease the $S_8$ tension~\cite{He:2023dbn,Dhyani:2025uof}. 
 To illustrate, we have considered SCDM with a mass of 0.1 MeV, which gives a $S_8$ central value of around 0.817.  In this case, the tension with the H0DN measurement of $H_0$ 
is reduced to 2.8$\sigma$. Further details, including the best fit values in the model, are given in the Table.~\ref{tab:additional_runs} in Appendix~\ref{sec:appendix}.

Finally, it is worth briefly commenting on how these models compare with extended frameworks, 
such as EDE and relevant extended versions in light 
of cosmological data. An updated analysis of EDE variants has recently appeared in Ref. \cite{Poulin:2025nfb}. The improved $H_0$ in \texttt{Model-B} yields \(H_0 = 69.83\pm0.98\ \mathrm{km\,s^{-1}\,Mpc^{-1}}\) for the 
\texttt{Planck\,+\,ACT DR6\,+\,DESI DR2} dataset combination, increasing to 
\(H_0 = 70.20\pm0.66\ \mathrm{km\,s^{-1}\,Mpc^{-1}}\) after including 
\texttt{SH0ES+Pantheon+}. 
These values are comparable to recent axion-like EDE analyses, where single-field EDE models typically prefer 
\(H_0 \sim 69.71\ \mathrm{km\,s^{-1}\,Mpc^{-1}}\) \citep{Poulin:2025nfb}, as obtained considering similar datasets. 
In two and three field EDE scenarios $H_0$ can increase further to \(\sim 70.5\text{--}70.9\ \mathrm{km\,s^{-1}\,Mpc^{-1}}\) ~\citep{Bella:2026zuk} with \texttt{Planck 2018 Commander} + \texttt{SimAll} + \texttt{Lensing(PR3)} + \texttt{CamSpec PR4} + \texttt{PantheonPlus} + \texttt{DESI DR2} datasets. 
Therefore, the behaviour of \texttt{Model-B} is broadly comparable with such resolutions of the Hubble tension. Also, recently, in Ref.~\cite{Jhaveri:2026bla} an EDE scenario combined with a late-time thawing DE model has been studied. Their inferred value of $H_0$ is consistent with the result obtained from \texttt{Model-B} in our analysis. 

\section{Conclusion}
\label{sec:conclusion}

In this article, a cosmological scenario is presented extending the $\Lambda$CDM with fractional DCDM ($f{\rm DCDM}$) and a DE component with an equation of state parameterized by $w_0$. The $f{\rm DCDM}$ component decays to massless DR species around the radiation--matter equality epoch. A comprehensive Bayesian MCMC analysis has been performed using \texttt{Planck~2018 TT/TE/EE}, \texttt{ACT~DR6}, \texttt{DESI~DR2 BAO}, \texttt{Planck CMB lensing}, and \texttt{SH0ES+Pantheon+} to estimate the relevant parameters. The important findings include the following.
\begin {itemize}
\item \texttt{Model-A} increases the Hubble constant to
$H_0 = 68.98 \pm 0.47\ \mathrm{km\,s^{-1}\,Mpc^{-1}}$
($68\%$ C.L.) by reducing $r_{\rm drag}$ through $f$DCDM decay around radiation--matter equality, accommodating a higher CMB-inferred $H_0$. \texttt{Model-B}, which additionally includes the DE equation-of-state parameter $w_0$, further raises it to
$H_0 = 69.83 \pm 0.98\ \mathrm{km\,s^{-1}\,Mpc^{-1}}$
($68\%$ C.L.), reducing the tension with \texttt{SH0ES} to $2.2\sigma$ and with \texttt{H0DN} to $2.9\sigma$. \texttt{Model-B} also provides an improved fit over $\Lambda$CDM with $\Delta \chi^2 = -8.8$ (full dataset), while remaining comparably favored under Bayesian model comparison.

\item The DE equation-of-state parameter ($w_0$) in \texttt{Model-B} marginally enters the phantom region. However, $w_0 = -1$ remains allowed at $1\sigma$. 

\item We observe that the constraint on $f_{\rm dcdm}$ remains nearly unchanged for both \texttt{Model-A} and \texttt{Model-B} when considering the \texttt{PLANCK+ACT (DR2)+lensing+DESI BAO (DR2)} dataset (serial no.1-4 in table.~\ref{tab:datasets}). However, upon adding the \texttt{SH0ES+Pantheon+} (serial no.5-6 in table.~\ref{tab:datasets})data, a non-zero central value for $f_{\rm dcdm}$ is obtained, with a preference for non-zero $f_{\rm dcdm}$ emerging at the $1\sigma$ level.

\item We also analyze an extension of \texttt{Model-B} including the summed neutrino mass parameter, $\sum m_\nu$. In this case, the upper bound on $\sum m_\nu$ weakens to $0.10\,\mathrm{eV}$ ($95\%$ C.L.), compared to upper limit on the $\Lambda$CDM$+\sum m_\nu$ bound of $0.068\,\mathrm{eV}$. Consequently, the model relaxes the neutrino mass constraints sufficiently to allow the possibility of an inverted neutrino mass hierarchy.

\item BBN and CMB constraints on $\Delta N_{\rm eff}$ are satisfied throughout the parameter space. For \(f_{\rm dcdm}\lesssim 0.06\), the modification to the expansion rate during BBN is negligible, while the decay occurs near matter--radiation equality, leading to only mild effects on the CMB. It is also important to note that \texttt{Model-B} achieves an improvement in $H_0$ comparable to that obtained in other extensions, such as Early Dark Energy (EDE) models, using the latest cosmological data.

\item The $S_8$ tension is not alleviated in \texttt{Model-B}; however, allowing the stable DM component (SCDM) to interact with electrons or protons can suppress small-scale power and reduce the $S_8$ tension, as illustrated.
\end{itemize}

Upcoming surveys such as \texttt{Euclid}~\cite{Euclid:2024pwi} and the Rubin Observatory \texttt{LSST}~\cite{LSSTDarkEnergyScience:2018yem} will significantly improve constraints on extended dark-sector scenarios through precise measurements of large-scale structure and cosmic expansion. These observations may provide decisive tests of DCDM frameworks and help clarify the origin of the $H_0$ tension.

\section*{Acknowledgements}
RD acknowledges CSIR, India for financial support through Senior Research Fellowship (File no. 09/ 1128 (13346)/ 2022 EMR-I) and Shiv Nadar IoE (Deemed to be University). AC acknowledges a helpful discussion with Subinoy Das. AAS acknowledges the funding from ANRF, Govt of India, under the research grant no. CRG/2023/003984. This article/publication is based upon work from COST Action CA21136- ``Addressing observational tensions in cosmology with systematics and fundamental physics (CosmoVerse)'', supported by COST (European Cooperation in Science and Technology). PM acknowledges funding from Anusandhan National Research Foundation (ANRF), Govt of India, Under the National Post-Doctoral Fellowship (File no. PDF/2023/001986).

\appendix

\section{}
\label{sec:appendix}

The following table depicts various relevant extended models in light of Sl.no 1-4 from Table.~\ref{tab:datasets} in Table.~\ref{tab:additional_runs}. The Bayesian evidence, together with $\chi^2$ analyses for \texttt{Model-B} have been presented in the Table.~\ref{tab:chi2_combined} 

\begin{table*}[htbp]
\centering
\footnotesize
\setlength{\tabcolsep}{3pt}
\caption{Mean values and $1\sigma$ uncertainties of cosmological parameters obtained from the combined \texttt{Planck 2018 + ACT + DESI DR2 (BAO)} data. The quantity $\Delta\chi^2 \equiv \chi^2_{\rm model} - \chi^2_{\Lambda\rm CDM}$ is computed with respect to the baseline $\Lambda$CDM model. Upper limits on $f_{\rm dcdm}$, $\sigma_{\rm dmeff}$, and $\Sigma m_\nu$ are at 95\% CL.}
\label{tab:additional_runs}
\resizebox{\textwidth}{!}{%
\begin{tabular}{lcccccc}
\hline\hline
Parameter
& $\Lambda$CDM
& $\Lambda$CDM$+\sum m_\nu$
& $w_0$CDM
& $w_0w_a$CDM
& $w_0w_a+f {\rm DCDM}$
& DM-$e$ Scattering \\
\hline
 $A_s\times10^{9}$
& $2.10\pm0.03$
& $2.130\pm0.024$
& $2.134\pm0.025$
& $2.11\pm0.003$
& $2.11\pm0.002$
& $2.134\pm0.023$ \\
$n_s$
& $0.974\pm0.003$
& $0.9744\pm0.0031$
& $0.9744\pm0.0034$
& $0.971\pm0.004$
& $0.975\pm0.004$
& $0.9784\pm0.0048$ \\
$100\,\theta_s$
& $1.0411\pm0.0003$
& $1.04179\pm0.00025$
& $1.04178\pm0.00025$
& $1.04169\pm0.00025$
& $1.04179\pm0.00026$
& $1.04199\pm0.00031$ \\
$\Omega_b h^2$
& $0.0224\pm0.00015$
& $0.02254\pm0.00010$
& $0.02254\pm0.00011$
& $0.02248\pm0.00011$
& $0.02246\pm0.00012$
& $0.02250\pm0.00012$ \\
$\Omega_{\rm cdm}h^2$
& $0.120\pm0.001$
& $0.1178\pm0.0006$
& $0.1176\pm0.0009$
& $0.1190\pm0.0011$
& $0.1228\pm0.0030$
& $0.1220\pm0.0034$ \\
$\tau_{\rm reio}$
& $0.0624\pm0.007$
& $0.0626\pm0.0061$
& $0.0609\pm0.0063$
& $0.059\pm0.006$
& $0.058\pm0.006$
& $0.0608\pm0.0063$ \\
$w_0$
& $-1$ (fixed)
& $-1$ (fixed)
& $-1.025\pm0.039$
& $-0.48\pm0.21$
& $-0.58\pm0.27$
& $-1.028\pm0.039$ \\
$w_a$
& -- 
& -- 
& -- 
& $-1.54\pm0.60$
& $-1.2\pm0.75$
& -- \\
$f_{\rm dcdm}$
& -- 
& -- 
& -- 
& -- 
& $<0.056$
& $<0.059$ \\
$\log_{10}\!\left(\dfrac{\Gamma_{\rm dcdm}}{\rm km\,s^{-1}\,Mpc^{-1}}\right)$
& -- 
& -- 
& -- 
& -- 
& $7.83\pm0.45$
& $7.72\pm0.52$ \\
$\sigma_{\rm dmeff}\,[{\rm cm}^2]$
& --
& --
& --
& --
& --
& $<3.35\times10^{-27}$ \\
$\sum m_\nu \,[{\rm eV}]$
& 0.06 (fixed) 
& $<0.068$
& 0.06 (fixed) 
& 0.06 (fixed) 
&  0.06 (fixed) 
& 0.06 (fixed)S \\
\hline
$\mathbf{H_0}$
& $\mathbf{68.38\pm0.28}$
& $\mathbf{68.59\pm0.29}$
& $\mathbf{69.11\pm1.00}$
& $\mathbf{64.1\pm1.9}$
& $\mathbf{65.4\pm2.52}$
& $\mathbf{69.69\pm1.09}$ \\
$\Omega_\Lambda (\Omega_{fld})$
& $0.690\pm0.04$
& $0.7011\pm0.0036$
& $0.7049\pm0.0081$
& $0.715 \pm0.005$
& $0.716 \pm 0.008$
& $0.705\pm0.008$ \\
$r_{\rm drag}$ [Mpc]
& $147.58\pm0.20$
& $147.51\pm0.19$
& $147.31\pm0.24$
& $147.24\pm0.26$
& $146.77\pm0.45$
& $146.95\pm0.48$ \\
$S_8$
& $0.817\pm0.007$
& --
& --
& --
& --
& $0.817\pm0.015$ \\
\hline
$\Delta\chi^2$
& $0$
& $-1.2$
& $-0.5$
& $-5.8$
& $-10.1$
& $-0.6$ \\
\hline\hline
\end{tabular}%
}
\end{table*}

\begin{table*}[t]
\centering
\small
\setlength{\tabcolsep}{4pt}
\renewcommand{\arraystretch}{1.15}

\caption{Best-fit $\chi^2$ values for $\Lambda$CDM and $w_0{+}f\mathrm{DCDM}$ for two dataset combinations.
$\Delta\chi^2$ is computed with respect to $\Lambda$CDM; negative values indicate improvement.
The Bayes factor $\ln\mathcal{B} = \ln(Z_{\Lambda\mathrm{CDM}}/Z_{w_0+f\mathrm{DCDM}})$. 
Dashes indicate datasets not included in a given combination.}
\label{tab:chi2_combined}

\resizebox{\textwidth}{!}{%
\begin{tabular}{l rrr rrr}
\toprule
& \multicolumn{3}{c}{(i) Planck+ACT DR6 + DESI DR2}
& \multicolumn{3}{c}{(ii) +SH0ES + Pantheon+} \\
\cmidrule(lr){2-4}\cmidrule(lr){5-7}
Dataset
& $\Lambda$CDM & \texttt{Model-B} & $\Delta\chi^2$
& $\Lambda$CDM & \texttt{Model-B} & $\Delta\chi^2$ \\
\midrule
\multicolumn{7}{l}{\textit{CMB}} \\
Planck low-$\ell$ TT
  & 21.72  & 21.57  & $-0.15$
  & 21.42  & 20.91  & $-0.51$ \\
Planck low-$\ell$ EE (sroll2)
  & 391.47 & 390.27 & $-1.20$
  & 390.35 & 390.46 & $+0.11$ \\
ACT DR6 CMB (PlanckActCut)
  & 221.19 & 221.98 & $+0.79$
  & 222.63 & 224.82 & $+2.19$ \\
ACT DR6 CMB
  & 159.20 & 155.36 & $-3.84$
  & 160.92 & 155.52 & $-5.40$ \\
ACT DR6 Lensing
  & 19.66  & 19.61  & $-0.05$
  & 20.18  & 19.92  & $-0.26$ \\
\textit{Total CMB}
  & 813.24 & 808.79 & $-4.45$
  & 815.48 & 811.61 & $-3.87$ \\
\addlinespace
\multicolumn{7}{l}{\textit{Large-scale structure \& distance}} \\
DESI BAO DR2
  & 11.10  & 11.89  & $+0.79$
  & 10.33  & 11.10  & $+0.77$ \\
SH0ES+Pantheon$+$
  & \multicolumn{3}{c}{---}
  & 1482.58 & 1476.91 & $-5.67$ \\
\midrule
\textbf{Total} $\chi^2$
  & \textbf{824.34} & \textbf{820.68} & $\mathbf{-3.66}$
  & \textbf{2308.42} & \textbf{2299.6268} & $\mathbf{-8.8}$ \\
\midrule
$\ln\mathcal{B}$
  & \multicolumn{2}{c}{}  & $+0.48 \pm 0.07$
  & \multicolumn{2}{c}{}  & $-0.9 \pm 0.08$ \\
\bottomrule
\end{tabular}
}
\end{table*}

\bibliography{references}
\end{document}